# Magnetic Monopole Noise

Ritika Dusad[1†], Franziska K.K. Kirschner[2†], Jesse C. Hoke[1,3], Benjamin Roberts[1], Anna Eyal[1,4], Felix Flicker[5], Graeme M. Luke[6,7,8], Stephen J. Blundell[2] and J.C. Séamus Davis[1,2,9]

1. Department of Physics, Cornell University, Ithaca, NY 14853, USA.
2. Clarendon Laboratory, Oxford University, Parks Road, Oxford, OX1 3PU, UK
3. Department of Physics, Stanford University, Stanford, CA 94305, USA
4. Department of Physics, Technion – Israel Institute of Technology, Haifa, 3200003, Israel
5. Rudolf Peierls Centre for Theoretical Physics, Clarendon Laboratory, Parks Road, Oxford, UK
6. Brockhouse Inst. for Materials Research, McMaster University, Hamilton, ON, Canada.
7. Department of Physics, McMaster University, Hamilton, Ontario, L8S 4M1, Canada.
8. Canadian Institute for Advanced Research, Toronto, Ontario, M5G 1Z8, Canada.
9. Department of Physics, University College Cork, Cork T12 R5C, Ireland.
† Contributed equally to this project.

**Magnetic monopoles [1-3] are hypothetical elementary particles exhibiting quantized magnetic charge $m_0 = \pm(h/\mu_0 e)$ and quantized magnetic flux $\Phi_0 = \pm h/e$. In principle, such a magnetic charge can be detected by the quantized jump in magnetic flux $\Phi$ it generates upon passage through a superconducting quantum interference device (SQUID)[4]. Naturally, with the theoretical discovery that a plasma of emergent magnetic charges should exist in several lanthanide-pyrochlore magnetic insulators[5,6] including $Dy_2Ti_2O_7$, this SQUID technique was proposed for their direct detection[6]. Experimentally, this has proven challenging because of the high number density, and the generation-recombination (GR) fluctuations, of the monopole plasma. Recently, however, theoretical advances have allowed the spectral density of magnetic-flux noise $S_\Phi(\omega, T)$ due to GR fluctuations of $\pm m_*$ magnetic charge pairs to be predicted [7,8]. Here we report development of a SQUID based flux-noise spectrometer, and consequent measurements of the frequency and temperature dependence of $S_\Phi(\omega, T)$ for $Dy_2Ti_2O_7$ samples. Virtually all the elements of $S_\Phi(\omega, T)$ predicted for a magnetic monopole plasma, including the existence of intense magnetization noise and its characteristic frequency and temperature dependence,** are detected. Moreover, comparisons of simulated and measured correlation functions $C_\Phi(t)$ of the magnetic-flux noise $\Phi(t)$ imply that the motion of magnetic charges is strongly correlated. Intriguingly, since the GR time constants $\tau(T)$ are in the millisecond range for $Dy_2Ti_2O_7$, magnetic monopole flux noise amplified by the SQUID is audible to human perception.

Observation of a quantized jump in magnetic flux $\Phi$ threading a SQUID loop when a monopole passes through it is the definitive technique for detection of magnetic monopoles[4] (Fig. 1A). It was proposed[6] for detection of thermally generated magnetic charges in magnetically-frustrated lanthanide-pyrochlore insulators[9,10] such as $Dy_2Ti_2O_7$ and $Ho_2Ti_2O_7$. Here, if each magnetic charge in a $\pm m_*$ pair departs to $\pm\infty$ in opposite directions, the net flux threading the SQUID loop should evolve from 0 to $\Phi_* = m_*\mu_0$ (Fig. 1B). However, because these materials are hypothesized to contain a dense plasma of equal numbers of $\pm m_*$ magnetic charges undergoing rapid thermal generation and recombination, $\Phi(t)$ measured by a SQUID (Fig. 1C) was expected to be stochastic and weak. Thus, despite extensive evidence for a magnetic charge plasma in $Dy_2Ti_2O_7$ and $Ho_2Ti_2O_7$ (Ref. 9,10), the magnetic-flux signature[6] of the magnetic charges $m_*$ has gone undetected.

In these compounds, the magnetic ions ($Dy^{3+}$;$Ho^{3+}$) occupy vertices of corner-sharing tetrahedra (Fig. 1D). Each ion exhibits only two spin configurations, as an Ising magnetic moment ($\mu \approx 10\mu_B$) that points either towards or away from the center of each tetrahedron[9,11] (black arrows Fig. 1D). The nearest-neighbor exchange interaction between these moments[9] takes the form $-J\sum \boldsymbol{S}_i \cdot \boldsymbol{S}_j$, with $J \approx -3.7K$ for $Dy_2Ti_2O_7$ and $J \approx -1.6K$ for $Ho_2Ti_2O_7$. The resulting dipolar spin ice (DSI) Hamiltonian incorporates both these exchange interactions and longer range dipole interactions as[12]

$$H = -J\sum_{<i,j>} \boldsymbol{S}_i \cdot \boldsymbol{S}_j + Da^3 \sum_{i<j} \left[ (\boldsymbol{S}_i \cdot \boldsymbol{S}_j)/|r_{ij}|^3 - \frac{3(\boldsymbol{S}_i \cdot \boldsymbol{r}_{ij})(\boldsymbol{S}_j \cdot \boldsymbol{r}_{ij})}{|r_{ij}|^5} \right] \qquad (1).$$

Here $D = \mu_0\mu^2/(4\pi a^3)$ is the dipole-interaction energy scale $D \approx +1.41K$ for both Dy$_2$Ti$_2$O$_7$ and Ho$_2$Ti$_2$O$_7$, and $a$=0.354 nm is the nearest-neighbor distance between moments. Only 6 energetically degenerate ground state spin configurations then occur per tetrahedron[13], all having two spins pointing in and two pointing out (Fig. 1D). The magnetic monopole model of the spin ices is achieved by re-writing the real Ising dipoles $\pm\mu$ of Eq. (1) in terms of magnetic charges $\pm m_*$ placed at the centers of adjacent tetrahedra such that $\pm\mu = \pm m_* d/2$ (Fig. 1D) where $d$ is the distance between tetrahedron centers $r_\alpha$. Each center $r_\alpha$ is then labeled by a net magnetic charge $m_\alpha$: $m_\alpha$=0 for 2-in/2-out, $m_\alpha = m_*$ for 3-out/1-in and $m_\alpha = -m_*$ for 3-in/1-out configurations (Fig. 1D). The interaction potential between such magnetic charges $m_\alpha$ and $m_\beta$ at sites $r_\alpha$ and $r_\beta$ is then represented by the Hamiltonian[6,9]

$$H = \frac{\mu_0}{4\pi}\sum_{\alpha<\beta}\frac{m_\alpha m_\beta}{r_{\alpha\beta}} + \frac{\nu}{2}\sum_\alpha m_\alpha^2 \qquad (2)$$

with $\nu = J/3 + 4/3\left(1+\sqrt{\frac{2}{3}}\right)D$. Here the first term represents the Coulomb-like interactions between magnetic charges, and the second an on-site repulsion which enforces the $m_\alpha$=0 or 2-in/2-out ground state at T=0. In this picture, when thermal fluctuations randomly flip a fraction of the Dy spins, this generates monopole quasiparticles with magnetic charges $\pm m_*$ plus a small additional population [14] with charge $\pm 2m_*$. Overall, the low-energy spin excitations are then hypothesized to be a plasma of $\pm m_*$ magnetic charges[5,6,9,10] interacting via a Coulomb-like potential, while undergoing rapid thermally activated generation and recombination across an energy barrier $\Delta \approx 2\nu(\mu/d)^2$.

Thermally generated plasmas of $\pm q$ electric charges, subject to both Coulomb interactions and spontaneous generation and recombination, are well understood in intrinsic semiconductors[15-16,17,18]. Here, thermal generation and recombination (GR) of $\pm q$ pairs generates a spectral density of voltage noise $S_V(\omega, T) = V^2 S_N(\omega, T)/N_0^2$, where $S_N(\omega, T)$ is the spectral density of GR fluctuations in the number of $\pm q$ pairs, and $N_0$ is their equilibrium density at a given temperature T. Analogous theories for GR fluctuations within a plasma of $\pm m_*$ magnetic-charge pairs have recently been developed for spin ice compounds[7]. The rates of $\pm m_*$ pair generation $g(N,T)$ and recombination $r(N,T)$ are such that, $g(N,T)|_{N_0} = r(N,T)|_{N_0}$ where $N_0(T)$ is the equilibrium number of magnetic charge pairs. However, thermally stimulated fluctuations $\delta N = N - N_0$ in the number of $\pm m_*$ pairs occur due to the generation and recombination processes. The Langevin Equation for these magnetic charge number fluctuations has been derived as[7]

$$\frac{d(\delta N)}{dt} = -\frac{\delta N}{\tau(T)} + \sqrt{A(T)}\zeta(t) \qquad (3)$$

where the GR rate is $\tau(T) = 1/(dr/dN - dg/dN)|_{N_0}$ and $\sqrt{A(T)}\zeta(t)$ represents the thermally generated stimulus uncorrelated in time[7,14]. Taking the Fourier transform of Eqn. 3 yields the predicted spectral density of $\pm m_*$ pair fluctuations as[7]

$$S_N(\omega, T) = \frac{4\sigma_N^2 \tau(T)}{(1 + \omega^2 \tau^2(T))} \qquad (4)$$

where $\sigma_N^2 \equiv \overline{(N - N_0)^2}$ is the variance in the number of $\pm m_*$ pairs. This result, shown schematically in Fig. 2A for a sequence of different $\tau(T)$, reveals many intriguing properties. The spectral density of $\pm m_*$ pair fluctuations should be constant up to an angular frequency $\omega_{GR}(T) \sim 1/\tau_{GR}(T)$ on the so-called GR noise plateau, which occurs because of the pure randomness of the GR processes. Above this frequency, $S_N(\omega, T)$ should eventually fall off as $1/\omega^2$ on time scales for which the uncorrelated magnetic charges are propagating freely. Another predicted signature of magnetic charge GR processes is that, in a regime where $\sigma_N^2$ is approximately constant[7], the power spectral density of lowest frequency number fluctuations $S_N(\omega \to 0, T)$ should increase proportionate to $\tau(T)$.

However, there remains a key challenge to SQUID detection of emergent magnetic monopoles[6]: the GR fluctuations of $\pm m_*$ pairs $S_N(\omega, T)$ must be related directly to fluctuations $S_\Phi(\omega, T)$ of magnetic flux $\Phi(t)$ detectable by a SQUID (Fig. 1C). Moreover, Eqn. 4 does not account for correlations in the motion of singly charged monopoles, nor for the existence of any doubly charged monopoles[14]. In theory, the correlations exist due to the topological constraints in spin ice that distinguish the dynamics of $\pm m_*$ pairs in these materials from free particle-antiparticle pairs. Monopoles of opposite sign are connected by Dirac strings with magnetic flux $\Phi_* = m_* \mu_0$ (yellow trace Fig 1D), whose existence forbids sequential traversal of the same trajectory by two magnetic charges of the same sign[10]. Therefore, a microscopic theory for magnetization fluctuations and noise, including the effects of both doubly-charged monopoles and correlated monopole motion, is required. Here we use Monte Carlo (MC) simulations of the thermally generated magnetic configurations from Eqn. 1 (Methods A), with the correspondence between the MC time-step and seconds established in Methods E. The result is a prediction of the spectral density of fluctuations $S_{M_z}(\omega, T)$ in the z-component of magnetization for each $\omega$ and $T$ in $Dy_2Ti_2O_7$. Figure. 2B shows such MC simulation-data presented as predicted noise spectral density of magnetic field fluctuations $S_{B_z}(\omega, T)$ for $Dy_2Ti_2O_7$ crystals of approximately our sample volume. Because our samples are $\approx 10^{16}$ larger in volume than can studied by MC simulations, the absolute magnitude of $S_{B_z}(\omega, T)$ is an estimation (Methods A), since effects of finite-size scaling on the DSI MC simulations[19] are unknown over such a volume range. Nevertheless, these MC simulations of Eqn. 1 contain several important predictions. First, the analytic form of $S_{B_z}(\omega, T)$ revealed by MC studies (Fig. 2B) is equivalent in key characteristics to Eqn. 4 (Fig. 2A) as derived using the Langevin Eqn. for GR fluctuations of $\pm m_*$ pairs. Second, when MC-simulation of $S_{B_z}(0, T)$ is plotted versus $\tau(T)$ for the temperature range of our experiments (Fig. 2C), they are approximately proportional, once an offset to all values of $S_{B_z}(0, T)$ due to numerical Nyquist (sampling) noise is considered. Third, by fitting the functional form $\tau(T)/(1 + (\omega \tau(T))^b)$ to the MC-simulation data in Fig. 2B, one can determine $\tau(T)$. Here the relationship $S(\omega) \propto \omega^{-2}$ at high frequencies from GR theory with a single GR time constant, is replaced with a more nuanced behavior $S(\omega) \propto \omega^{-b}$ with $b(T) < 2$, possibly due to a combination of spectra as in Eqn. 4 with a distribution of GR time constants. In any case, as discussed below, the MC simulations indicate that this $S(\omega) \propto \omega^{-b}$ behavior in $Dy_2Ti_2O_7$ derives microscopically from the topological constraints from the Dirac strings causing the motion of monopoles to become correlated.

These theoretical innovations provide clear predictors for the magnetic-flux noise spectral density $S_\Phi(\omega, T)$ due to a plasma of $\pm m_*$ magnetic charges undergoing GR processes. First, random fluctuations dominate at frequencies $\omega\tau(T) \ll 1$ and topologically constrained monopole dynamics dominate at $\omega\tau(T) \gg 1$, with a transition regime surrounding $\omega\tau(T) \approx 1$. Second, even though the evolution with $T$ of magnetic charge recombination times $\tau(T)$ is microscopically complex[10], the GR prediction for a plasma of $\pm m_*$ magnetic charges is that

$$S_\Phi(0, T) \propto \tau(T) \qquad (5)$$

provided that $\sigma_N^2$ remains approximately constant in the temperature range of the experiment (Ref. 7 and Methods B). Third, the signatures of topological constraints on the interactions between monopoles are contained within the prediction[8] for the magnetic-flux noise autocorrelation function $C_{B_Z}(t)$ and power-law $b(T)$. These concepts are embodied in predictions that

$$S_\Phi(\omega, T) \propto \tau(T)/[1 + \left(\omega\tau(T)\right)^b] \qquad (6)$$

To exploit these opportunities, we developed a high-sensitivity SQUID based flux-noise spectrometer[20,21] to measure $S_\Phi(\omega)$ generated by crystalline samples. Here we use it for determination of $S_\Phi(\omega, T)$ from $Dy_2Ti_2O_7$ samples over a frequency range 1Hz<$f$<2.5kHz (Methods C). This spin-noise spectrometer (SNS) is mounted on a custom-built, low-vibration, variable temperature cryostat operable in the range 1.2K≤T≤7K, which was estimated to be optimal for detection of most intense noise spectra from the millimeter scale $Dy_2Ti_2O_7$ crystals studied. Measurements then consist of varying the temperature of the SQUID and sample assembly, from 1.2K to 4K in steps of 25mK, and using a spectrum analyzer to measure the $S_\Phi(\omega, T)$ generated by $Dy_2Ti_2O_7$ samples throughout this temperature range.

Figure 3A shows a typical example of $S_\Phi(\omega, T)$ measured from $Dy_2Ti_2O_7$ samples in our SNS for $1.2K \leq T \leq 4K$. The left-hand axis is the magnetic-flux noise spectral density $S_\Phi(\omega, T)$ while the right-hand axis shows an estimate of the equivalent magnetic-field noise spectral density $S_{B_Z}(\omega, T)$ within the $Dy_2Ti_2O_7$ samples (Methods C). Each data set is fit to Eq. (6) with the best fit shown as a fine solid curve (Methods D). Thus, we find that the magnetic-flux noise spectral density of $Dy_2Ti_2O_7$ is constant for frequencies from near 1Hz up to an angular frequency $\omega(T) \sim 1/\tau(T)$, above which it falls off as $\omega^{-b}$ where $b$ spans a range between 1.2 and 1.5. Figure 3B shows $S_\Phi(\omega, T)/S_\Phi(0, T)$, revealing that the GR time constant $\tau(T)$ evolves rapidly toward longer times at lower T. Figure 3C shows the measured $S_\Phi(0, T)$ from 3A plotted against the measured $\tau(T)$ from fits in 3A, where T is the implicit variable ranging from 1.2K to 4K. Thus, $S_\Phi(0, T)$ of $Dy_2Ti_2O_7$ is proportional to $\tau(T)$ throughout our temperature range (Fig. 2C; Fig. 4A). This situation is as expected within GR models when the variance in monopole number $\sigma_N^2$ remains roughly constant, as confirmed in Methods D. Comparison of GR time constants $\tau(T)$ derived from flux-noise spectroscopy to the relaxation times $\tau_M(T)$ derived from magnetic susceptibility measurements [22-25] reveal them to have good empirical correspondence (Methods D), although their microscopic relationship remains to be identified. Finally, since the monopole times $\tau(T)$ are in the millisecond range, magnetic monopole flux noise amplified by the SQUID is actually audible to human perception (audio files attached). The theoretical predictions for a thermally generated plasma of $\pm m_*$ magnetic charges dominated by generation-recombination fluctuations (Fig. 2) are consistent with all of these unusual magnetic-flux noise phenomena (Fig. 3, 4A).

The frequency dependence of $S_\Phi(\omega, T)$ contains additional key information. Monte-Carlo simulations for $Dy_2Ti_2O_7$ predict the autocorrelation function $C_{B_Z}(t,T)$ of magnetic field fluctuations $B_Z(t)$ (Ref. 8). Figure 4B shows $log\left[C_{B_Z}(t,T)/C_{B_Z}(0,T)\right]$ predictions for three distinct magnetic charge dynamics theories at T=1.2K. The first MC model (blue) describes $\pm m_*$ magnetic charge plasma of dipolar spin ice (DSI), which has Coulomb-like interactions and Dirac-strings[8,10] (yellow Fig. 1D). The second MC model (green) is the nearest neighbor spin ice model (NNSI) in which Coulomb-like interactions are suppressed but Dirac-string constraints present. The final model (red) is a neutral plasma of $\pm m_*$ magnetic charges undergoing Coulomb interactions that is topologically unconstrained. For comparison, our measured autocorrelation function $log\left[C_{B_Z}(t,T)/C_{B_Z}(0,T)\right]$ is plotted in black with a best-fit curve overlaid. Clearly, the DSI model, including Coulomb-like interactions and Dirac-string topological constraints, is far more consistent with the directly measured correlation function in this system. Moreover, the NNSI model is inconsistent with the experiment as short time correlations appear to be completely absent. This correlation function phenomenology is virtually unchanged, except for values of $\tau(T)$, within our temperature range. Equivalently, the MC predictions of frequency exponent $b$ from the same three theories DSI (blue), NNSI (green) and free monopoles (red) can be determined by fitting each simulated $S_{B_Z}(\omega, T)$ to $\tau(T)/\left(1+(\omega\tau(T))^{b(T)}\right)$. The results are shown in Fig. 4C with the measured $b(T)$ from fitting to $S_{B_Z}(\omega, T)$ in Fig. 3A shown as black dots. We note that the GR theory in Eqn. 4 is well supported by these MC simulations, which predict $S_\Phi(\omega, T) \propto \tau/(1+(\omega\tau)^b)$ where $b$=2 for uncorrelated free monopoles and $b$<2 for the full DSI Hamiltonian. Moreover, comparison between simulated and measured autocorrelation functions $C_{B_Z}(t,T)$ and falloff exponents $b$, for magnetic-flux noise in $Dy_2Ti_2O_7$ reveals that the DSI model is most consistent with the observed phenomenology. To achieve precise agreement may require adjustment of the $J, D$ terms in Eqn. 1(Ref. 26), or better control over finite size scaling effects[19]. Overall, however, Fig. 4C implies that the power-law signatures of strong correlations observed in both $log\left[C_{B_Z}(t,T)/C_{B_Z}(0,T)\right]$ and $S_{B_Z}(\omega, t)$ are occurring due to a combination of the existence of a Dirac-string trailing each monopole (Fig. 1D) and the Coulombic interactions.

To recapitulate: theoretical predictions for the magnetic-flux signature of a $\pm m_*$ magnetic charge plasma[6,7,8] in spin ice, are studied for the case of $Dy_2Ti_2O_7$. Strong magnetization noise as predicted by Monte-Carlo simulations[8] (Fig. 2) is observed for the first time (Fig. 3A). The frequency and temperature dependence of the magnetic-flux noise spectrum $S_\Phi(\omega, T)$ predicted for $\pm m_*$ magnetic charges undergoing thermal generation and recombination (Fig. 2B,C) is confirmed directly and in detail (Fig. 3). The expected transition from a plateau of constant magnetic-flux noise[7,8] for $\omega\tau(T) \ll 1$, to a power-law falloff[8] for $\omega\tau(T) \gg 1$, is observed throughout (Fig. 3A). And the low-frequency flux-noise spectral density increases rapidly as $S_\Phi(0,T) \propto \tau(T)$ with falling $T$ (Fig. 4A). These $S_\Phi(\omega, T)$ characteristics are exceptional, in that the magnetization-noise spectral density signature of a ferromagnet[20], a classic spin glass[21], or an Ising paramagnet[27] all evolve qualitatively differently with $\omega$ and $T$. On the other hand, within the context of $\pm m_*$ generation-recombination theory, the observed $S_\Phi(\omega, T)$ (Figs. 3,4) is quite consistent with other studies which imply that $Dy_2Ti_2O_7$ and $Ho_2Ti_2O_7$ contain a plasma of emergent magnetic monopoles[9,10,14,28-33]. Additionally, the agreement of measured magnetization noise autocorrelation functions $C_{B_z}(t)/C_{B_z}(0)$ with those predicted from MC simulations (Fig. 4B), implies significant correlations in the motions of $Dy_2Ti_2O_7$ magnetic charges. Overall, we find detailed and comprehensive agreement between current theories for thermal generation and recombination of a correlated $\pm m_*$ magnetic monopole plasma (Fig. 2) and the phenomenology of magnetic-flux noise spectral density in $Dy_2Ti_2O_7$ (Figs. 3,4).

# FIGURES

**Figure 1**

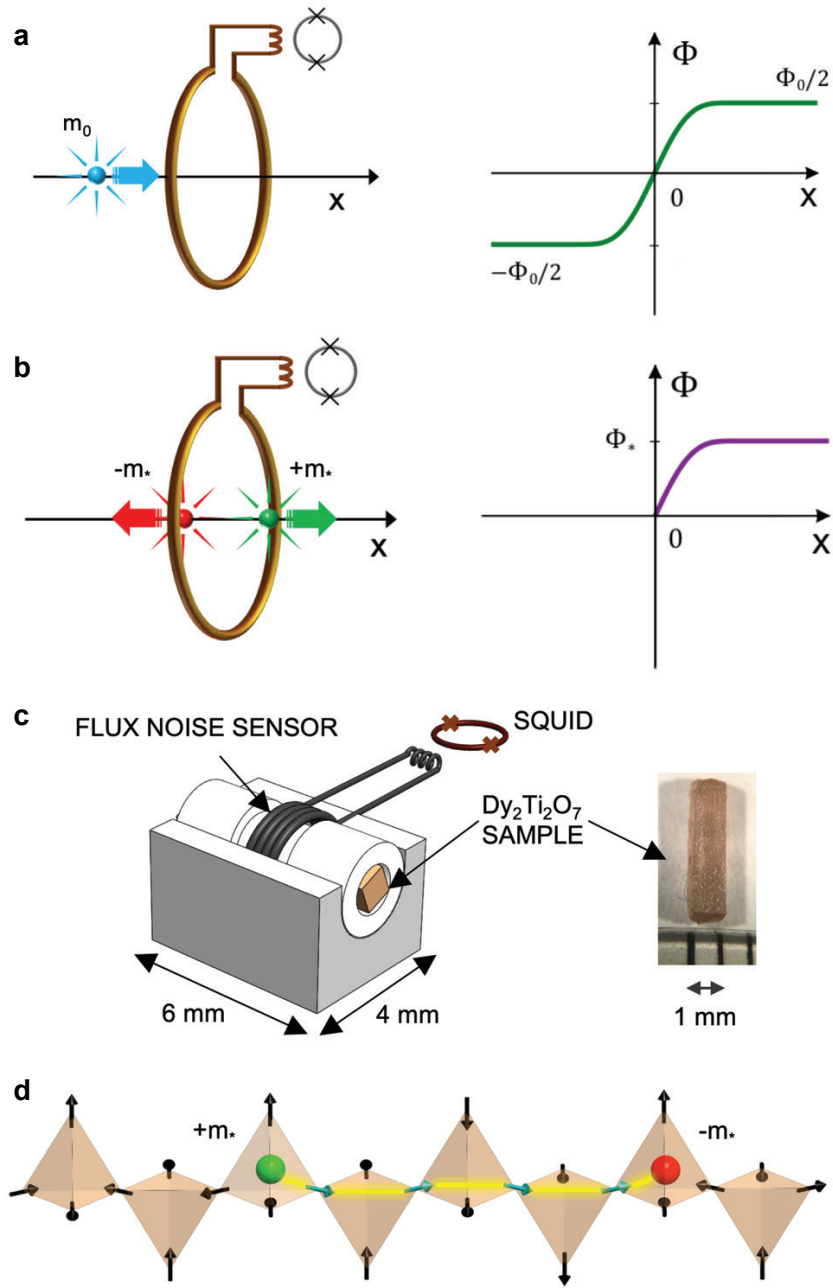

**Figure 1 SQUID-based Quantized Flux Jump Detection of Magnetic Monopole**

A. Schematic of fundamental Dirac magnetic monopole with charge $m_0$ traversing, from $x = -\infty$ to $x = +\infty$, through the input-coil of the SQUID. The magnetic-flux threading the SQUID changes in total by $\Phi_0 = h/e$.

B. Schematic of two emergent magnetic charges $\pm m_*$ generated in $Dy_2Ti_2O_7$ at x=0 by a thermal fluctuation. As each charge departs in opposite directions to x= $\pm\infty$, the net flux threading the SQUID changes in total by $\Phi_* = \mu_0 m_*$.

C. Schematic of the spin-noise spectrometer used throughout this project (Methods C). The primary coil, consisting of six turns of 0.09mm diameter NbTi wire wound on the cylindrical MACOR shell, and its two connecting wires to the input coil of a Quantum Design 550 DC SQUID, are contained within a cylindrical superconducting Nb flux shield (not shown). The thermal conductivity of MACOR is sufficient to thermalize the sample for all T>1K. Inset: single crystals of $Dy_2Ti_2O_7$ cut in the shape of a square-cross-sectional rod (photo) are inserted along the axis of the pickup coil.

D. Schematic representation of the spin ice excited state in which two magnetic charges $\pm m_*$ are generated by a spin flip, and propagate through the material. In $Dy_2Ti_2O_7$ the tetrahedron corners are the midpoints of the bonds of a diamond lattice defined by the centers $r_\alpha$ of the tetrahedra, and all such bonds point along the local [111] direction. The ratio of the lattice constant of the diamond $d$ and fundamental pyrochlore lattice $a$ is $d = \sqrt{3/2}a$. A single flip of an Ising $Dy^{3+}$ spin converts the 2-in/2-out $m_\alpha = 0$ configuration in adjacent tetrahedra, to a situation with adjacent $m_\alpha = m_*$ for 3-out/1-in in one and $m_\alpha = -m_*$ for 3-in/1-out in the next. These two opposite magnetic charges can then separate via a sequence of spin flips in sequential tetrahedra which leave them all in the 2-in/2-out $m_\alpha = 0$ configuration except at the ends of this chain. However, due to the spin ice constraints the specific path taken, which can be identified with a Dirac string with flux $\Phi_*$, cannot be traversed sequentially by another such magnetic charge of the same sign.

**Figure 2**

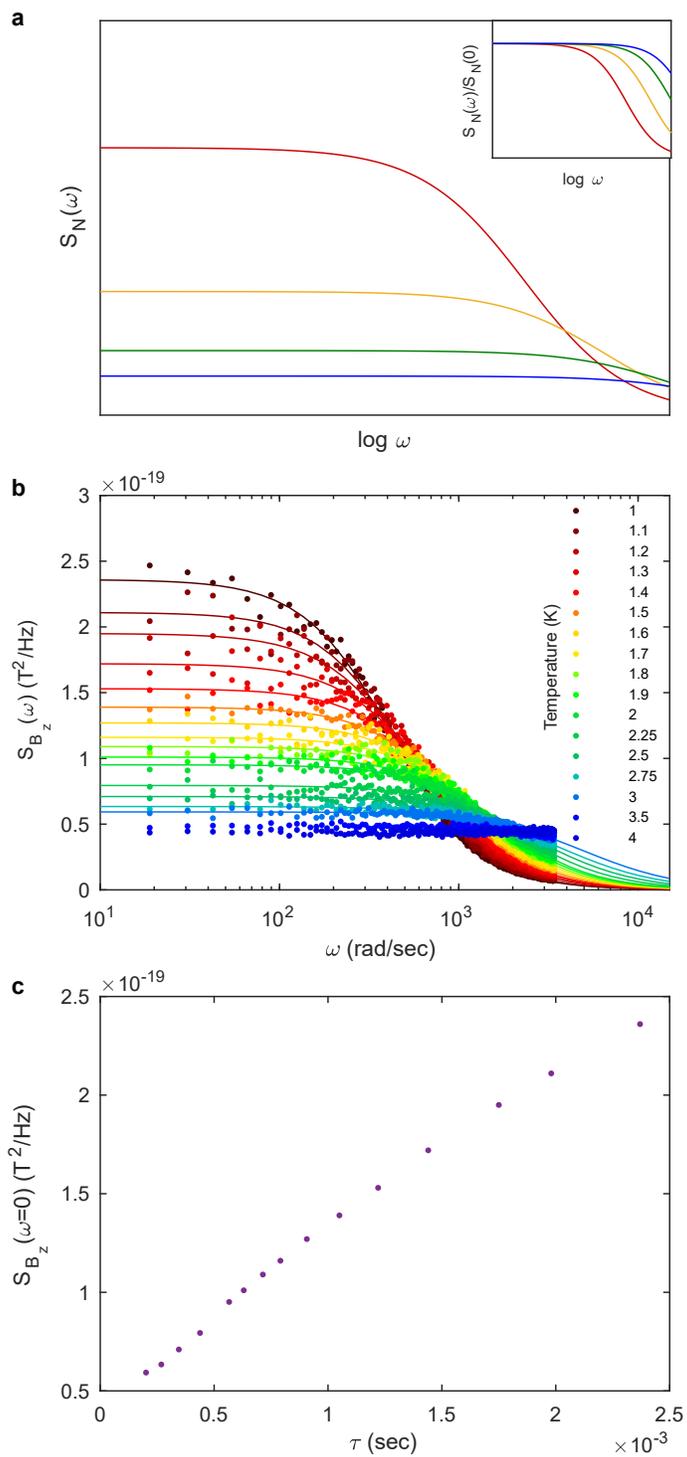

**Figure 2 Spectral Density of Fluctuations in Monopole Number and Magnetization**

A. Predicted spectral density of fluctuations in monopole number $S_N(\omega, T)$ from Eqn. 4, for several monopole GR time constants $\tau$, in a range that one might expect to achieve by cooling $Dy_2Ti_2O_7$ from 4K to ~1K. The GR plateau in $S_N(\omega, T)$ is clear as $\omega \to 0$ as is the $\omega^{-2}$ falloff expected of free monopole motion for frequencies $\omega\tau > 1$. Top right corner inset shows normalized spectral density $S_N(\omega, T)/S_N(0, T)$.

B. Predicted spectral density of magnetic field fluctuations within the $Dy_2Ti_2O_7$ sample $S_{B_z}(\omega, T)$ from MC simulations using the Hamiltonian in Eqn. 1 in a range 1K≤T≤4K. These simulations are physically quite distinct from those reported in Ref. 8 because here we focus on bulk fluctuations of the z-component of magnetization $M_z(t)$ or of the associated z-component of magnetic field $B_z(t)$, while Ref. 8 considers field fluctuations in vacuum outside a specific crystal termination surface. Because of periodic boundary conditions and the very small MC sample volume, scaling of the predicted magnitude of $S_{B_z}(\omega, T)$ for the MC sample, to the absolute magnitude expected for an experimental sample of volume ~ $10^{16}$ greater is an approximation (Methods A). But this does not affect the form of $S_{B_z}(\omega, T)$ expected for a macroscopic experimental sample.

C. Predicted relationship from $Dy_2Ti_2O_7$ MC simulations[8] of $B_z(t)$, of $S_{B_z}(0, T)$ versus $\tau(T)$ for the GR fluctuations of a $\pm m_*$ magnetic charge plasma. Note that all $S_{B_z}(0, T)$ are offset by a constant along the y-axis due to artifacts of Nyquist (sampling) noise at the high frequency end of the MC calculations

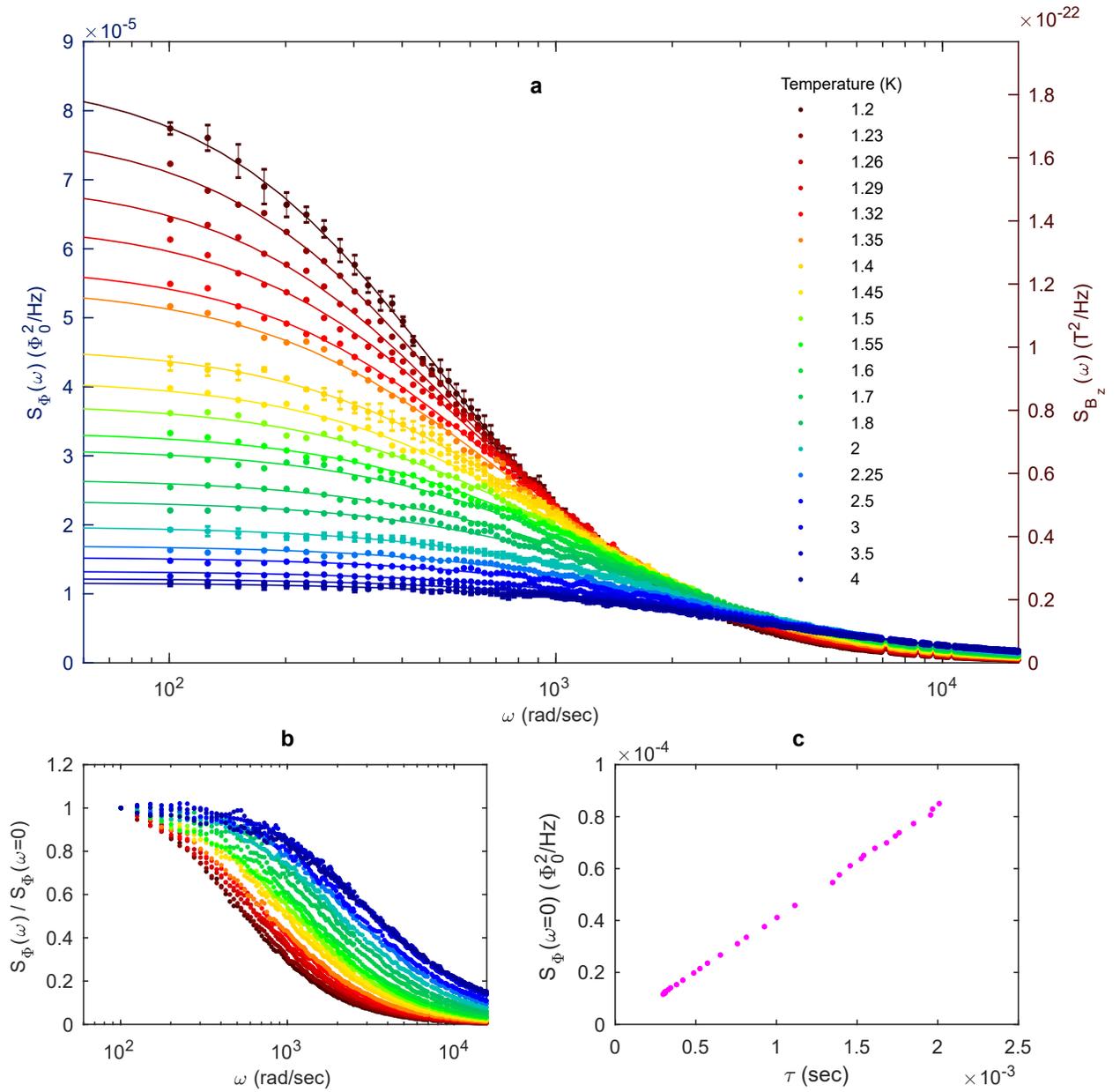

**Figure 3 Spectral Density of Magnetic-Flux Noise in Dy$_2$Ti$_2$O$_7$**

A. Measured spectral density of flux-noise $S_\Phi(\omega, T)$ from Dy$_2$Ti$_2$O$_7$ samples (e.g. Fig. 1C) in the range 1.2K≤T≤4K. The left-hand axis is the magnetic-flux noise spectral density $S_\Phi(\omega, T)$; the right-hand axis is an estimate of the equivalent magnetic-field noise spectral density $S_{B_Z}(\omega, T)$ averaged over the Dy$_2$Ti$_2$O$_7$ samples (based on calibration of the flux sensitivity of the spectrometer Methods C). The best fit to the function $\tau(T)/\left(1 + (\omega\tau(T))^{b(T)}\right)$ shown as a fine solid curve. Overall we find $S_\Phi(\omega, T)$ of Dy$_2$Ti$_2$O$_7$ to be constant for frequencies 1Hz < $(T) = 1/2\pi\tau(T)$ , above which it falls off as $\omega^b$. Error bars are shown for four temperatures (1.20K, 1.40K, 2.00K, 4.00K) for visual clarity. Magnitude of error bars represent the standard deviation of each data point extracted from an average of five independent Dy$_2$Ti$_2$O$_7$ flux-noise datasets, at each temperature.

B. Normalized spectral density of flux noise $S_\Phi(\omega, T)/S_\Phi(0, T)$ revealing the divergence of the time constant $\tau(T)$ toward longer times at lower T (SI Fig. 5)

C. $S_\Phi(0, T)$ plotted versus $\tau(T)$ as measured from fitting data in 3A. Observation that $S_\Phi(0, T) \propto \tau(T)$ for Dy$_2$Ti$_2$O$_7$ throughout the full temperature range is a key expectation for $\pm m_*$ GR magnetic-flux noise.

**Figure 4**

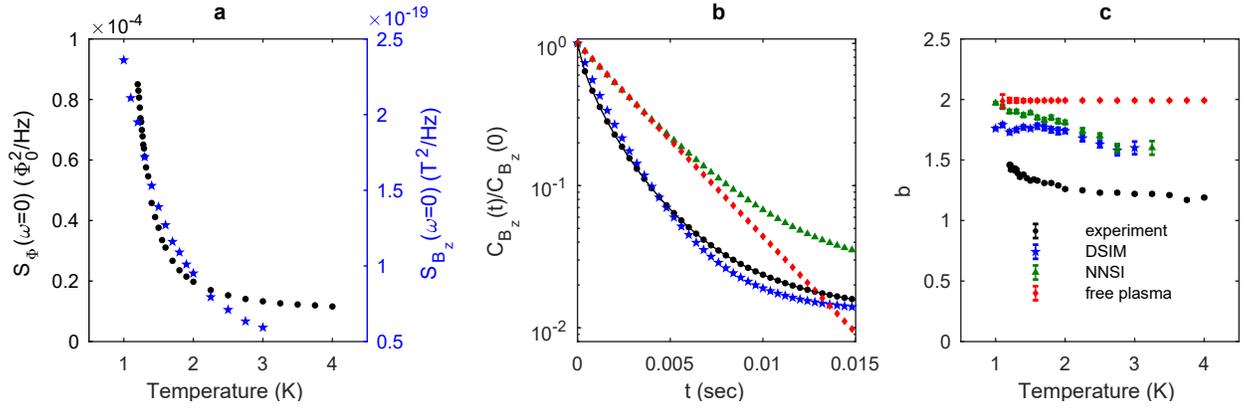

**Figure 4 Magnetic Flux Noise Spectral Density of Correlated Monopole Fluid**

A.  Experimentally measured $S_\Phi(0,T)$ (black) and MC predicted $S_{B_z}(0,T)$ (blue) low frequency noise spectral density of Dy$_2$Ti$_2$O$_7$ with falling temperature.

B.  For T=1.2K, the Monte-Carlo simulation prediction of the autocorrelation function $log\left[\frac{C_{B_z}(t)}{C_{B_z}(0)}\right]$ in the field fluctuations $B_Z(t)$ for three models of the spin dynamics of Dy$_2$Ti$_2$O$_7$. The DSI model contains Coulomb-like interactions and constraints on repeated passage of a same-sign monopoles along the same trajectory due to Dirac strings (blue); the nearest neighbor spin ice model (NNSI) has had the Coulomb interactions suppressed (green); the free monopole plasma (red) is in-keeping with free monopole GR theory. The measured autocorrelation function $log\left[\frac{C_{B_z}(t)}{C_{B_z}(0)}\right]$ of magnetic-field fluctuations $B_Z(t)$ of Dy$_2$Ti$_2$O$_7$ is plotted in black and overlaid with a fit function; the experimental error bars are smaller than the data points.  Clearly the autocorrelation function of the DSI model corresponds best to the measured $C_{B_z}(t)$. We note that the distinction between the single slope (red) for the free monopole plasma[7], and the more complex predicted $C_{B_z}(t)$ for the other cases, represents microscopically a distinction between a simple process involving a single time constant vs. a more complex one, potentially involving a spread of relaxation time constants. Most importantly, the measured $C_{B_Z}(t)$ (black) shows that magnetization dynamics is obviously strongly correlated in time.

C. Monte-Carlo prediction of exponent $b$ in $S_{B_Z}(\omega, T) \propto \tau(T)/\left(1 + (\omega\tau(T))^{b(T)}\right)$ for the three magnetic charge dynamics theories. As in B, these are the DSI model (blue); the NNSI model (green); the free plasma model (red). Measured exponent $b$ from fitting $S_{\Phi}(\omega, T) \propto \tau(T)/\left(1 + (\omega\tau(T))^{b(T)}\right)$ for all data in Fig. 3A is shown in black. The time constants $\tau(T)$ of the MC $S_{B_Z}(\omega, T)$ and of the $S_{\Phi}(\omega, T)$ data are not free parameters here (Fig 4A). Error bars represent the uncertainty in analytical fits to data due to statistical fluctuations of $S_{\Phi}(\omega, T)$ data.

## References

1   Dirac, P. Quantised Singularities in the Electromagnetic Field. *Proceedings of the Royal Society A: Mathematical, Physical and Engineering Sciences* 133, 60-72 (1931)

2   Hooft, G. Magnetic monopoles in unified gauge theories. *Nuclear Physics B* 79, 276-284 (1974)

3   Polyakov, A.M. Particle Spectrum in the Quantum Field Theory, *JETP Lett.* 20, 194 (1974).

4   Cabrera, B. First Results from a Superconductive Detector for Moving Magnetic Monopoles, *Phys. Rev. Lett.* 48, 1378 (1982)

5   Rhyzkin I. A. Magnetic relaxation of rare-earth pyrochlores, *JETP 101*, 481 (2005)

6   Castelnovo C., Moessner R., Sondhi S.L. Magnetic monopoles in spin ice, *Nature 451*, 42 (2008)

7   Klyuev, A., Ryzhkin, M. & Yakimov, A. Statistics of Fluctuations of Magnetic Monopole Concentration in Spin Ice. *Fluctuation and Noise Letters* 16, 1750035 (2017)

8   Kirschner, F.K.K., Flicker, F., Yacoby, A., Yao, N. & Blundell, S.J. Proposal for the detection of magnetic monopoles in spin ice via nanoscale magnetometry. *Physical Review B* 97, (2018)

9   Castelnovo, C., Moessner, R. & Sondhi, S. Spin Ice, Fractionalization, and Topological Order. *Annual Review of Condensed Matter Physics* 3, 35-55 (2012)

10   Jaubert, L. & Holdsworth, P. Magnetic monopole dynamics in spin ice. Journal of Physics: Condensed Matter 23, 164222 (2011)

11   Rosenkranz S. *et al* Crystal-field interaction in the pyrochlore magnet $Ho_2Ti_2O_7$. *Journal of Applied Physics* 87:5914-5916. (2000)

12   den Hertog B., Gingras M Dipolar Interactions and the Origin of Spin Ice in Ising Pyrochlore Magnets. *Phys. Rev. Lett.* 84: 3430. (2000)

13  Ramirez, A., Hayashi, A., Cava, R., Siddharthan, R. & Shastry, B. Zero-point entropy in 'spin ice'. *Nature* 399, 333-335 (1999)

14  Kaiser V., *et al*, Emergent electrochemistry in spin ice: Debye-Hückel theory and beyond Phys. Rev. B 98, 144413 (2018)

15  Burgess, R. E. The Statistics of Charge Carrier Fluctuations in Semiconductors. *Proc. Phys. Soc. B* 69 1020 (1956)

16  van Vliet, K. M. and Fassett, J. R. Fluctuation Phenomena in Solids, edited by Burgess, R. E. *Academic Press, New York* (1965)

17  Mitin V., Reggiani L., and Varani L. *Noise and Fluctuations Control in Electronic Devices*, Chapter 2, American Scientific Publishers (2002)

18  Konczakowska A. and Wilamowski B.M. *Fundamentals of Industrial Electronics*, Chapter 11, Taylor & Francis (2011)

19  Melko, R. G. and Gingras, M. J. P. Monte Carlo studies of the dipolar spin ice model *J. Phys.: Condens. Matter* 16 R1277 (2004)

20  Vitale, S., Cavalleri, A., Cerdonio, M., Maraner, A. *and* Prodi, G.A., Thermal equilibrium noise with $1/f$ spectrum in a ferromagnetic alloy: Anomalous temperature dependence, *Journal of Applied Physics* 76, 6332 (1998).

21  Reim, W., Koch, R., Malozemoff, A., Ketchen, M. & Maletta, H. Magnetic Equilibrium Noise in Spin-Glasses:Eu0.4Sr0.6S. *Physical Review Letters* 57, 905-908 (1986)

22  Snyder, J. et al. Low-temperature spin freezing in the Dy2Ti2O7 spin ice. *Phys. Rev. B.* 69 064414 (2004)

23  Matsuhira, K. et al. Spin dynamics at very low temperature in spin ice Dy2Ti2O7. *J. Phys. Soc. Jpn.* 80 123711 (2011)

24  Yaraskavitch et al. Spin dynamics in the frozen state of the dipolar spin ice material $Dy_2Ti_2O_7$. Phys. Rev. B. 85, 020410 (2012)

25  Kassner, E. R.  et al. Supercooled spin liquid state in the frustrated pyrochlore $Dy_2Ti_2O_7$. *Proc. Nat. Acad. Sci.* 112 8549 (2015)

26  Fennell, T.  et al. Neutron scattering investigation of the spin ice state in $Dy_2Ti_2O_7$. Phys. Rev. B. 70 134408 (2004)

27  Quilliam, J. A., Meng, S., Mugford, C. G. A., Kycia, J. B., Evidence of Spin Glass Dynamics in Dilute LiHoxY1−xF4, *Phys. Rev. Lett.* 101, 187204 (2008)

28  Morris D *et al.* Dirac Strings and Magnetic Monopoles in the Spin Ice Dy2Ti2O7. *Science* **326**: 411-414 (2009)

29  Bramwell S. T. *et al* , Measurement of the charge and current of magnetic monopoles in spin ice, *Nature*  **461**,  956–959 (2009)

30  Bovo L, Bloxsom J, Prabhakaran D, Aeppli G, Bramwell S Brownian motion and quantum dynamics of magnetic monopoles in spin ice. *Nat. Comm.* 4:1535-1542. (2013)

31   Giblin S, Bramwell S, Holdsworth P, Prabhakaran D, Terry I Creation and measurement of long-lived magnetic monopole currents in spin ice. *Nat. Phys.*, 7:252-258. (2011)

32  Kaiser, V., Bramwell, S. T., Holdsworth, P. C. W., Moessner, R. ac Wien Effect in Spin Ice, Manifest in Nonlinear, Nonequilibrium Susceptibility. Phys. Rev. Lett. 115 037201 (2015)

33  Paulsen, C. et al. Experimental signature of the attractive Coulomb force between positive and negative magnetic monopoles in spin ice. Nat. Phys. 12 661 (2016)

**Acknowledgements:** We are grateful to C. Castelnovo, J. Goff, Yong Beak Kim, M.J. Lawler, A. Ramirez, D. Schlom and N. Y. Yao for very helpful discussions and communications. J.C.S.D. thanks Owen H.S. Davis for insightful discussions and for proposing to study magnetic noise in pyrochlores. R.D. thanks Y.X. Chong for extensive aid during experimental operations, and acknowledges use of Cornell Center for Materials Research Shared Facilities supported through the NSF MRSEC program (DMR-1719875). A.E. acknowledges support from the Israeli Pazy Equipment Grant 299/18. F.K.K.K. acknowledges support from Lincoln College, Oxford. F.F. acknowledges support from the Astor Junior Research Fellowship of New College, Oxford. Conceptual design of the experimental technology was supported by the W.M. Keck Foundation. J.C.S.D. acknowledges support from Science Foundation Ireland under Award SFI 17/RP/5445 and from the European Research Council (ERC) under Award DLV-788932. R.D. and J.C.S.D acknowledge support, plus funding for instrument development and the experimental studies, from the Gordon and Betty Moore Foundation's EPiQS Initiative through Grant GBMF4544.

**Author Contributions:** R.D. and J.C.S.D. conceptualized the project and designed the experimental setup. R.D. developed the flux noise spectrometer. R.D., J.C.H., B.R. and A.E. carried out the experiments and data analysis. G.M.L. synthesized the sequence of $Dy_2Ti_2O_7$ samples. F.K.K.K. carried out the Monte Carlo simulations with help from F.F. G.M.L., S.J.B. and J.C.S.D. supervised the investigation and wrote the paper with key contributions from R.D., F.K.K.K. and F.F. The manuscript reflects the contributions of all authors.

**Author Information:** Reprints and permissions information is available at www.nature.com/reprints. The authors declare no competing financial interests. Readers are welcome to comment on the online version of the paper. Correspondence and requests for materials should be addressed to S.J.B.; stephen.blundell@physics.ox.ac.uk or to J.C.S.D.; jcseamusdavis@gmail.com.

**Methods**

## *A    Monte Carlo Simulations*

### A.1.    *MC Simulation Procedures*

Monte Carlo (MC) simulations are used to model the magnetization dynamics of $Dy_2Ti_2O_7$. In general, these simulations were carried out on a sample containing 4x4x4 unit cells, each of which contains 16 $Dy^{3+}$ ions. We refer to this as the MC sample. Standard MC procedures are used[34] consisting of $10^6$ cooling steps followed by an interval of 5000 MC-time-steps, at fixed T. During this interval W, the time dependence of net z-component of magnetic moment $\mu_Z(t)$ of the whole MC sample is then simulated. This procedure is then repeated 600 times. Carrying out the MC simulations for 5000 MC sequential time-steps ensures capture of a spread of microscopic time-scales at each temperature which covers that expected in this material at these temperatures. This is because it spans the range from approximately a single MC step ~100uS to approximately the total time for which simulation runs ~1s. The range of temperatures of these simulations was between 4.0K and 1 K. Because this is a simulation of bulk magnetization dynamics, periodic boundary conditions were used in all directions.

For a given conformation, the z-component of magnetic moment of the MC sample $\mu_Z$ was found by summing z-components over the individual magnetic moments of the 1024 Dy spins, where $\mu \approx 10 \mu_B$. The simulated time dependence of this value at a given temperature T is $\mu_Z(t, T)$, and is evaluated sequentially during the time window W. Its autocorrelation function is

$$C_{\mu_Z}(\tau, T) = \frac{1}{W} \int_{-W/2}^{W/2} \mu_Z(t, T) \, \mu_Z(t + \tau, T) dt \quad [\mu_B^2\,] \qquad \text{(M.1)}$$

The predicted spectral density of magnetic moment noise in the MC sample is then calculated using the Wiener-Khinchin theorem

$$S_{\mu_Z}(\omega, T) = 4 \int_0^\infty C_{\mu_Z}(\tau, T) cos(\omega\tau) d\tau \quad [\mu_B^2\, MCstep] \qquad \text{(M.2)}.$$

We extract the frequency range $10^{-4}$( MC-steps )$^{-1}$ to $10^{-1}$( MC-steps )$^{-1}$ (the Nyquist frequency is 0.5(MC steps)$^{-1}$). Equation M.2 was then averaged over the 600-independent simulation runs to the get better precision for $S_{\mu_Z}(\omega, T)$ . Finally, the time axis of the MC simulation is converted from MC-step to seconds as described in Methods E, so that angular frequency in Eqn. M.2 and throughout the paper is $\omega(\frac{rad}{sec}) = 2\pi/t(sec)$ and $S_{\mu_Z}(\omega, T)$ $[\mu_B^2\, s]$.

To convert to the noise spectral density of the z-component of the magnetization, we use $S_{M_Z}(\omega, T) = \frac{\mu_B^2 S_{\mu_Z}(\omega,T)}{V^2}$ $[A^2 m^{-2} s]$ where V=6.6X10$^{-26}$ m$^3$ is the volume of the MC sample. To estimate the predicted spectral density of z-magnetization noise expected from the Dy$_2$Ti$_2$O$_7$ experimental sample within range of the SQUID pickup coil (Fig. 1C), we estimate that there are N= $2.9 \cdot 10^{16} \pm 20\%$ MC samples in the experimental volume, and divide $S_{M_Z}(\omega, T)$ by N. The spectral density of fluctuations of z-component of magnetic field B[Tesla] within the sample is then

$$S_{B_Z}(\omega, T) = \mu_o^2 S_{M_Z}(\omega, T) \quad [T^2 s] \tag{M.3}.$$

Figure 2B then shows this estimation of $S_{B_Z}(\omega, T)$ from the DSI Hamiltonian (Eqn. 1) for our specific sample geometry in the spin noise spectrometer.

### A.2 *Model Hamiltonians*

First the full dipolar spin ice Hamiltonian (given in Eq. (1) in the main text) is employed. The exchange energy is J ≈ −3.72 K and the dipolar energy is D ≈ 1.41 K for DTO. This dipolar spin ice model (DSIM) leads to a lowest energy state of Dy spins pointing into the center or out of the tetrahedron they are sitting on along the local <111> axes. This state is known as the 2-in-2-out state (Fig. 1D). As previously discussed, the violation of this rule by a spin flip causes generation of a monopole anti-monopole pair with charge $\pm m_*$ or doubly charged pair with charge ±2m* (Ref. 14). The energy of two nearest-neighbor monopoles is 3.06K, and the energy to create one monopole is $\Delta$ = 4.35 K. Since the spins sit on tetrahedral corners, magnetic monopole motion is guided by spin flips in a topologically constrained fashion. These monopoles experience a strong coulombic force between the $\pm m_*$ charges as described in Eqn. (2) of the main text.

Second the nearest-neighbor spin ice (NNSI) Hamiltonian is considered. This is executed by setting D=0 in Eqn. 1 of main text. It suppresses the effects of long-range coulombic interactions. *J* is chosen such that the system still has a 2-in/2-out ground state, while having the same density of excitations as DSI at a given temperature. This system still has monopole-like excitations, but greatly reduced force between the monopoles. Thus, we predict the noise spectral density of $\pm m_*$ magnetic charge pairs hopping in the presence of strongly suppressed coulombic interactions.

Third, we identify the noise spectrum of $\pm m_*$ magnetic charge pairs hopping freely in the absence of Coulomb interactions or topological constraints due to the Dirac strings in Dy$_2$Ti$_2$O$_7$. The model is specified in Eq. (2) in the main text, with $m_*$ charges located on the sites of a diamond lattice.

## B    Statistics of monopole number fluctuations

The master equation for generation $g(N,T)$ and recombination $r(N,T)$) of magnetic monopole pairs [Ref .7] defines the probability $P(N,T)$ of finding $N$ monopole pairs at temperature $T$ in a steady state condition

$$\frac{dP(N,T)}{dt} = r(N+1,T)P(N+1,T) + g(N-1,T)P(N-1,T)$$
$$- P(N,T)[g(N,T) + r(N,T)] = 0 \qquad (M.4)$$

Here $g(N,T)$ and $r(N,T)$ generation and recombination rates of the monopoles, and add or removed one magnetic charge pair per generation or recombination event, respectively. The exact dependence of $g$ and $r$ on $N$ and $T$ depends on the microscopics of generation and recombination process pertaining to the specific system under investigation. Absent fluctuations, we also note that (Ref. 15,16)

$$\frac{d}{dt}\langle \delta N\rangle = -\langle \delta N\rangle \frac{d(r-g)}{dN}\Big|_{N=N_0} \qquad (M.5)$$

where $\delta N = N - N_0$ . The time constant for N to approach its equilibrium value is defined

$$\frac{1}{\tau(T)} \equiv \frac{d(r-g)}{dN}\Big|_{N_0} = r'(N_0,T) - g'(N_0,T) \qquad (M.6)$$

Obviously, in equilibrium, $g(N_0,T)= r(N_0,T)$.

Expanding $\ln P(N,T)$ about its maximum value $\ln P(N_0,T)$ in a quadratic fashion (Ref. 15,16) yields

$$\left[\frac{\partial^2}{\partial N^2}\ln P(N,T)\right]_{N=N_0} = \frac{g'(N_0,T)}{g(N_0,T)} - \frac{r'(N_0,T)}{r(N_0,T)} \qquad (M.7)$$

so that

$$\ln P(N,T) = \ln P(N_0,T) - \frac{1}{2}(N-N_0)^2\left[\frac{r'(N_0,T)}{r(N_0,T)} - \frac{g'(N_0,T)}{g(N_0,T)}\right] \qquad (M.8)$$

Thus the expected Gaussian probability distribution of N about its most probable value $N_0$ is

$$P(N,T) = P(N_0,T)\exp\left[-(N-N_0)^2/\overline{2(N-N_0)^2}\right] \qquad (M.9)$$

The variance of monopole number $\sigma_N^2 = \overline{(N-N_0)^2}$ is then determined from equations M.7,9 (Ref. 15,16) as

$$\sigma_N^2(T) = \left[\frac{r'(N_0,T)}{r(N_0,T)} - \frac{g'(N_0,T)}{g(N_0,T)}\right]^{-1} = \frac{g(N_0,T)}{r'(N_0,T)-g'(N_0,T)} = g(N_0,T)\cdot \tau(T) \qquad (M.10)$$

For emergent magnetic monopoles in $Dy_2Ti_2O_7$, the equilibrium generation rate at temperature T within our range will approximately be $g(N_0,T) \propto \exp(-\Delta/T)$ where $\Delta$ *is* the thermal energy barrier for thermal spin flips that generate the monopoles. It is established from previous experiments that at these higher temperatures the time constants are given approximately by $\tau(T)\sim\exp(\Delta/T)$ (Ref. 23). This implies that the variance of magnetic monopole number $\sigma_N^2 \sim \exp\left(\frac{\Delta}{T}\right) \cdot \exp\left(-\frac{\Delta}{T}\right)$ would be expected to be approximately constant in this temperature range.

### C    Spin-noise Spectrometer

#### C.1 Design

We use a custom-built 1K cryostat to carry out our experiments, with the SQUID and the sample-holder (SH) plus their superconducting shielding mounted below and thermalized to the 1K pot refrigerator. The spin noise spectrometer (SNS) setup consists of cylindrical SH with a concentric hole of diameter 1.4 mm and length 5.7 mm is used to encapsulate rod-shaped samples as shown in Fig. 1C. The superconducting pickup coil is wound around the SH and consists of 6 turns of thin Nb wire with inductance $L \approx 0.25\ \mu H$. The input inductance of the QD 550 SQUID is reported by the manufacturer (Quantum Design) to be $=2.0\ \mu H$. The SH and the SQUID circuitry of the SNS are all contained within a Nb tube of aspect ratio R~4 for flux shielding. No external cables enter this shielded region, to minimize external noise being picked-up by our detector. This insures that flux-noise floor of this apparatus sits at minimum level quoted by manufacturer $\Phi < 4\ \mu\phi_0\ /\sqrt{Hz}$ for the entire temperature range of study as shown in E.D. Figure 1.

#### C.2. Operation

A typical operation cycle consists of cooling down experiment to 1.2K and then using PID controls to vary temperature from 1.2K to above 4K with temperature stability at each point of 2.5 mK. Once the temperature is stable at a desired set-point, we record the flux noise generated by our sample via a spectrum analyzer. Unprocessed data at each temperature consists of 5 datasets of BW 2.5kHz, each of which is an outcome of averaging of 1000 acquisitions where acquisition time = (1/resolution BW), resulting in the measured SQUID output as voltage noise spectral density detected at the SQUID $S_v(\omega, T)$.

#### C.3 Calibration

The transfer function C between pickup coil and SQUID is calibrated by driving a small known flux $\Phi_{TEST}(\phi_0)$ via a drive coil (inserted into the pickup coil) through the pickup coil, and recording the corresponding SQUID output voltage $V_S$. In this case (E.D. Figure 1)-

$$C = \left(\frac{V_S}{0.684}\right)\frac{1}{\Phi_{TEST}(\phi_0)} \tag{M.11}.$$

We find that C=0.015. The spectral density of magnetic-flux noise within the sample is obtained

$$S_\Phi(\omega, T) = S_v(\omega, T)/(C^2). \tag{M.12}$$

To relate $S_\Phi(\omega, T)$ to the magnetic field noise spectral density generated within our sample $S_{Bz}(\omega, T)$, we consider the cross-sectional area of the sample $\sigma = 1.4 X 10^{-6} m^2 \pm 17\%$ yielding

$$S_{Bz}(\omega, T) = S_\Phi(\omega, T)/(\sigma)^2 \ [\text{T}^2\text{s}] \tag{M.13}.$$

### C.4 *Flux-Noise Signal Strength*

A typical magnetic-flux noise spectral density from a sample compared to the noise spectral density of an empty pickup coil is shown in E.D. Figure 2 for BW: 1Hz to 2.5kHz. The plateau of flux-noise spectral density from $Dy_2Ti_2O_7$ sits a factor of $1.5 \times 10^6$ higher than the noise floor level. Mechanical noise peaks for empty coil flux signal and for $Dy_2Ti_2O_7$ flux signal have been deleted manually. We note that the plateau for flux-noise spectral density signal of monopoles from $Dy_2Ti_2O_7$ goes down to at least 1Hz.

### C.5 *Sample Geometry Effects*

Shape effects could occur in such spin noise measurements. This is because even though we are measuring a cuboidal sample with a coil around the middle, spins at the ends of the sample still contribute partially. Our experiments measure flux through the pickup coil due to the dipole fields from spins in the sample, and the noise is coming fundamentally from spin flips (aka monopole hops). The flux at the pickup coil due to a single spin in the sample depends on where that spin is and which way it is pointing. Some spins are invisible (e.g. spins pointing to a direction in the xy plane) and some have a greater effect, e.g. spins close to, but not at the edge, in or near the plane of the coil, pointing along z. If there is a fluctuation of the magnetic moment of the whole sample, resulting in a net moment, then that would produce a net demagnetization field which all the spins would experience, potentially affecting their dynamics, and that demagnetization field would be shape-dependent.

## D.    Data Analysis

### D.1 *Fitting*

The flux-noise spectral density floor measured for empty pickup coil in the SNS, i.e. background noise $S_\Phi^{bcg}$, is fitted to a smooth polynomial function $S_\Phi^f$. Since the noise floor does not vary with temperature, the same function $S_\Phi^f$ is then subtracted from the measured $Dy_2Ti_2O_7$, $S_\Phi^{DTO}(\omega, T)$ for all temperatures, to obtain the reported noise spectral density $S_\Phi(\omega, T)$ dataset shown in Fig. 3.

This post processed data is then fit to empirical equation (6) using Least Squares method for a BW of 16Hz-2.5kHz for all temperatures. While the plateau in flux-noise spectral density $S_\Phi(\omega, T)$ goes down to at least 1Hz for all temperatures, to optimize data acquisition times to $\sim 1$ hour per temperature, for all spectra reported in Fig. 3, the lower limit for BW of data for regression analysis is set at 16Hz. The time constant $\tau(T)$, power law for frequency $b(T)$ and $S_\Phi(0, T)$ are free parameters in the fitting procedure and fits for all temperatures are of high quality with $R^2 > 0.99$. The residuals for these fits are shown in E.D. Figure 3.

We established that the flux noise spectral density coming from $Dy_2Ti_2O_7$ is repeatable in different single crystals of the material as shown, for a typical example, in E.D. Figure 4.

D.2 *Comparison of time constants*

It has been established empirically that the susceptibility-derived time constants $\tau_M(T)$ representing the magnetization dynamics of $Dy_2Ti_2O_7$ diverge with decreasing T (Refs 22,23,24) and are likely heterogeneous[25,35]. This type of ac susceptibility measurement has been made for $Dy_2Ti_2O_7$ in different sample shapes ranging from polycrystalline samples[24] to toroidal single crystals[25,35]. Yaraskavitch *et al* established that the time constants $\tau_M(T)$ measured from SQUID-based susceptibility measurements of rod shaped samples were in good qualitative agreement with $\tau_M(T)$ reported in both Snyder *et al* and Matsuhira *et al*. Eyvazov *et al* (Ref. 35) verified quantitative agreement between $\tau_M(T)$ from ac susceptibility measurements and $\tau_M(T)$ from Yaraskavitch *et al*.

In E.D. Fig. 5, we show the GR time constants $\tau(T)$ obtained from our flux-noise experiments. These clearly follow a quantitatively equivalent trajectory to the ac susceptibility $\tau_M(T)$ of Ref. 25, 35 and thus correspond well with the $\tau_M(T)$ derived from numerous susceptibility studies of this material (Refs. 22,23,24,25). However, the correspondence between $\tau_M(T)$ and $\tau(T)$ reported in this work remains to be understood within a quantitative microscopic theory.

Additionally, evidence for a heterogeneous distribution of microscopic spin relaxation rates contributing to $\tau_M(T)$ had been adduced from the stretched-exponential form for the time dependence of magnetization[25, 35] in $Dy_2Ti_2O_7$. From consideration of the related ac susceptibility studies[22-25,35] and the similarity between $\tau_M(T)$ and $\tau(T)$, we estimate that any effects of such heterogeneity on the flux-noise spectrum $S_\Phi(\omega, T)$ would first become visible at temperatures below T=1K. Therefore, they are unlikely to have been detected in these first SNS studies.

D.3 *Variance in magnetic-flux noise*

We note that the measured variance of the signal $\Phi(t)$, $\sigma_\Phi^2$, is approximately constant as a function of temperature as shown in E.D. Fig. 6. This variance is determined at each temperature by integrating measured flux noise spectrum over frequency.

$$\sigma_\Phi^2 = \int_0^\infty S_\Phi(\omega)d\omega \qquad (M.\ 14)$$

The result in E.D. Fig. 6 may appear surprising because one would generically expect that the fluctuations should be rapidly suppressed with falling temperatures. However, although the number of magnetic monopoles in $Dy_2Ti_2O_7$ is believed to decrease rapidly (exponentially in the simplest theory) with falling temperature, the integrated magnetic-flux noise power in the temperature range of our studies remains approximately constant. As explained in Methods B, this phenomenology is empirically consistent with measured $\tau(T)$. However, it will require further theoretical and experimental studies to determine and understand its evolution towards lower temperatures.

## E    Calibration of MC Time Step

E.1 *Inter-calibration of time scales*

To obtain a correspondence between MC-step time and actual time, we assume that the MC temperature is equal to the measured temperature of the experiment in the range 1.2K to 3K. We then compare generation-recombination time constants obtained from fitting $S(\omega,T)$ to $\tau(T)/(1 + (\omega\tau(T))^b$ for a. MC DSIM simulations in Fig. 2B to yield $\tau_{MC\ DSIM}(T)$, and b. experimental data in in Fig. 3A to yield $\tau_{experiment}(T)$, respectively. The temperature T is used as the implicit variable. The slope of a linear fit of $\tau_{MC\ DSIM}(T)$ versus $\tau_{experiment}(T)$ yields the correspondence between MC-step and actual time. The result yields that in these simulations 1MC-step = 83$\pm$ 11 microseconds which is then used throughout our studies. The possibility of nonlinearity in the correspondence of our MC-step to seconds, for example as reported in other MC simulations (Takatsu *et al,* JPSJ 82, 104710 (2013)), will require future studies on how scaling the periodic boundary conditions and MC sample volumes impacts the linear correspondence that is typically used in similar experiment-MC comparisons (Ref. 10).

**METHODS REFERENCES:**

34.  Metropolis, A. W. Rosenbluth, M. N. Rosenbluth, A. H. Teller, and E. Teller, J. Chem. Phys. 21, 1087 (1953)),

35. Eyvazov, A. B. et al. Common glass-forming spin liquid state in the pyrochlore magnets $Dy_2Ti_2O_7$ and $Ho_2Ti_2O_7$. *Phys. Rev. B.* 98 214430 (2018)

# Extended Data Figures

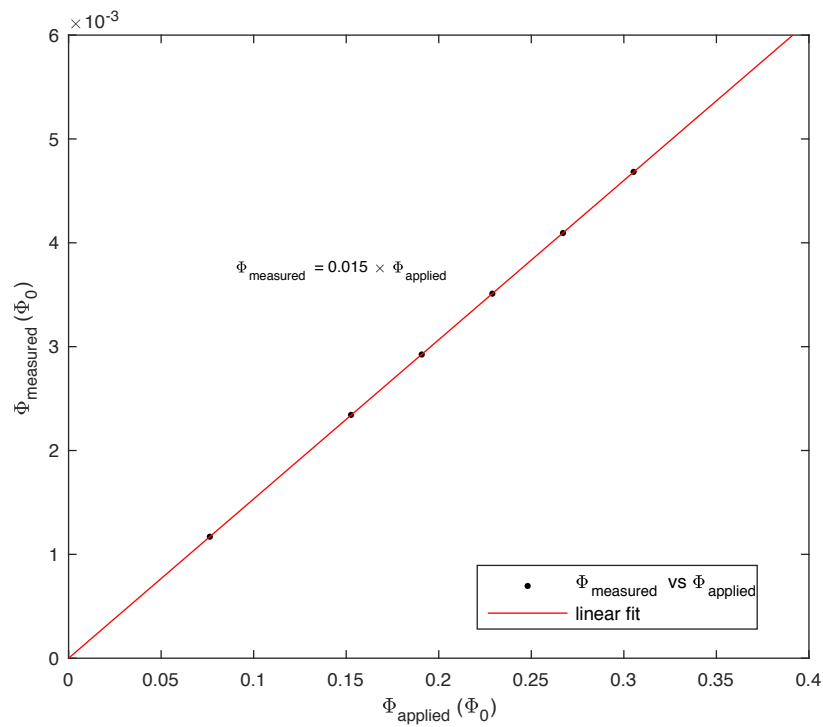

**Extended Data Figure 1:** Here we show linear relationship between flux applied to the pickup coil via a drive coil and flux output by the SQUID. The slope between the two gives us the transfer function between the pickup coil and the SQUID.

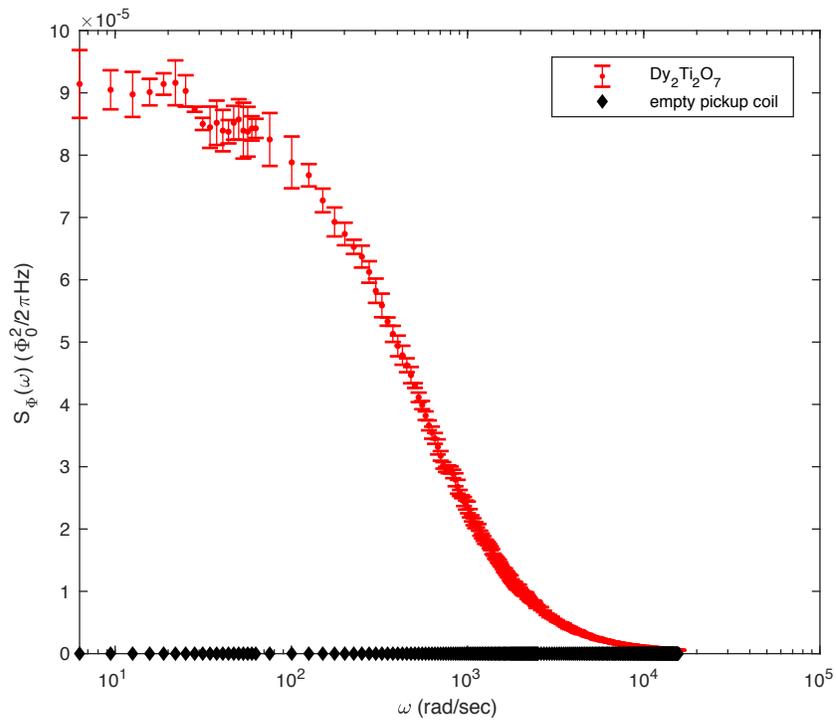

**Extended Data Figure 2**: A typical spectrum of magnetic-flux noise spectral density detected from a sample of Dy$_2$Ti$_2$O$_7$ (at 1.22K) compared to flux-noise spectral density of empty pickup coil corresponding to $\sim 16.8 \times 10^{-12}$ $\phi_0^2$ /Hz. The black data points are rendered meaningfully visible by a vertical shift in the position of 0.

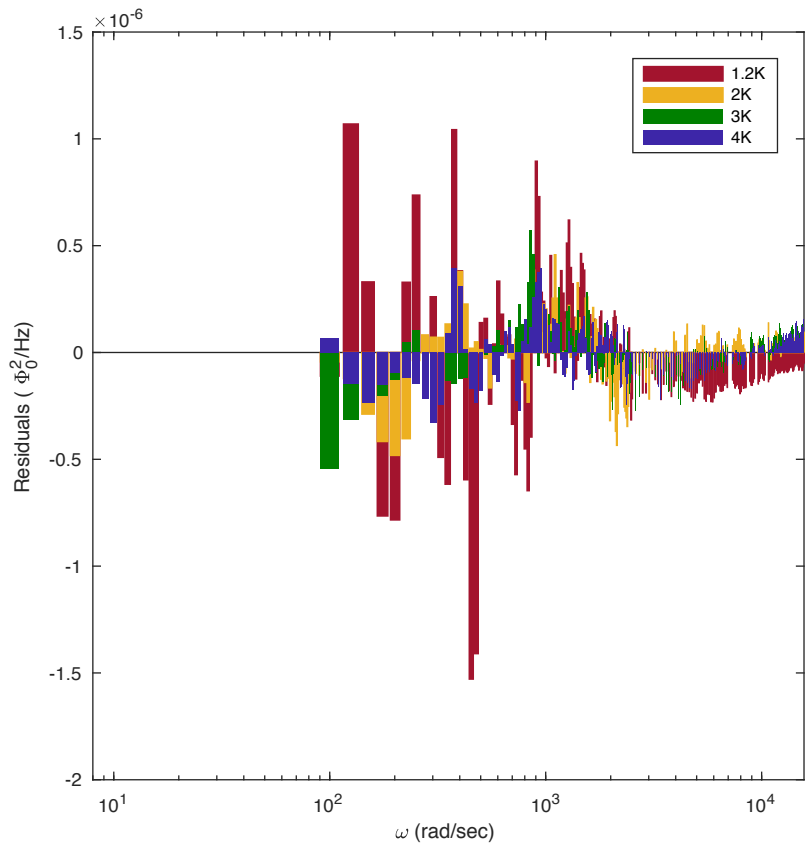

**Extended Data Figure 3**: Residuals $(S_\Phi(\omega, T) - S_{FIT}(\omega, T))$ for fits of measured flux noise spectral density (Fig. 3) to Eq. (5) are shown here for four temperatures.

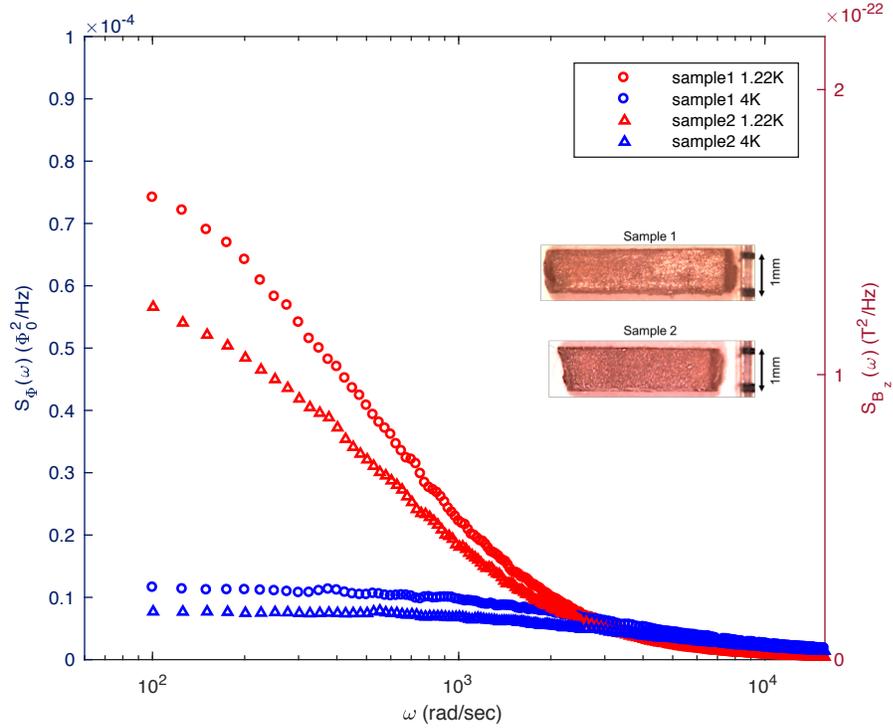

**Extended Data Figure 4**: Plot of magnetic flux noise $S_\Phi(\omega, T)$ from two different Dy$_2$Ti$_2$O$_7$ rod shaped samples. We observe that $S_\Phi(\omega, T)$ from these two typical samples is very similar and therefore this experiment is qualitatively quite repeatable on single crystals of Dy$_2$Ti$_2$O$_7$. The differences in magnitude and time constant are due to the geometrical differences between the two samples.

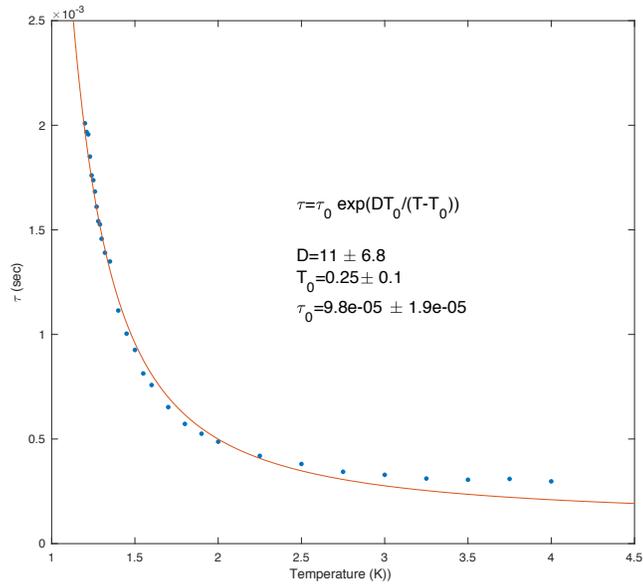

**Extended Data Figure 5:** Plot of time constant from fits to measured $S_\Phi(\omega, T)$ data as shown in Fig. 3. The flux-noise derived time constant $\tau(T)$ behaves in a super Arrhenius fashion $\tau(T) = \tau_0 \exp\left(D\frac{T_0}{T-T_0}\right)$, consistent with previous measurements of ac susceptibility time constants $\tau_M(T)$ (MR 1) which also exhibited super-Arrhenius time constant behavior with $\tau_0 \sim 1.4 \times 10^{-4}$ s, D $\approx 14$, $T_0 \approx 0.26$ K.

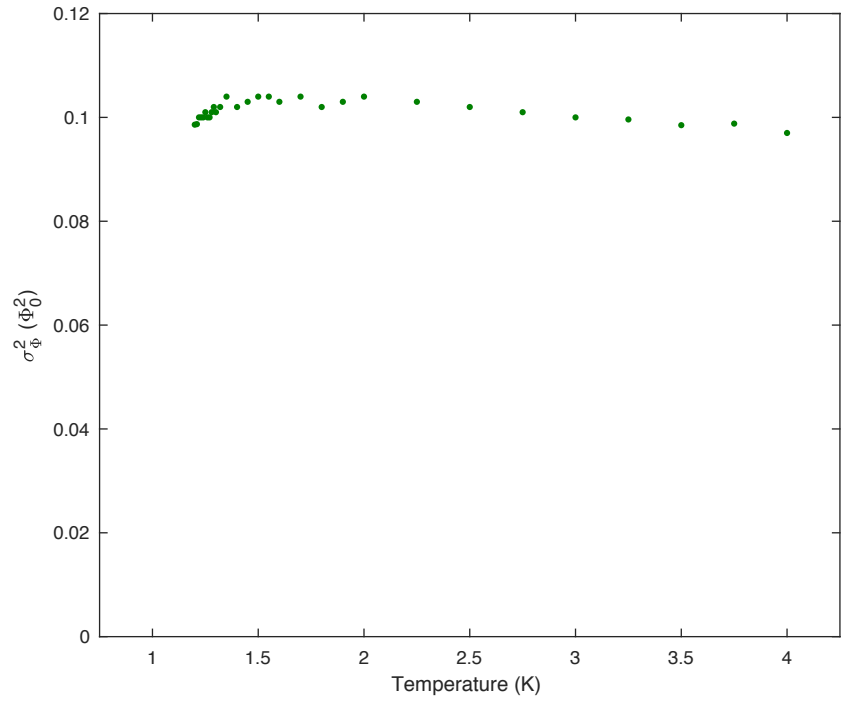

**Extended Data Figure 6**: Plot of measured variance of flux $\sigma^2_\Phi$ shows that it is approximately constant as a function of temperature, in this range.